# Total Routhian surface calculations for neutron-rich $^{149}$Ce*

R. Kaczarowski and W. Płóciennik

The Andrzej Sołtan Institute for Nuclear Studies, PL-05-400 Świerk, Poland

and

A. Syntfeld[1], H. Mach[2], W. Kurcewicz[1], B. Fogelberg[2] and P. Hoff[3]

[1]Institute of Experimental Physics, Warsaw University, Hoża 69, 00-681 Warsaw, Poland
[2]Department of Radiation Sciences, Uppsala University, 61 182 Nyköping, Sweden
[3]Department of Chemistry, University of Oslo, P.O. Box 1033, Blindern, N-0315 Oslo, Norway

A total Routhian surface (TRS) calculations were performed for the $^{149}$Ce nucleus. The equilibrium deformation parameters and total angular momentum values have been calculated as a function of rotational frequency for both signatures of the lowest positive- and negative-parity rotational bands. Theoretical predictions are compared with the results of experimental studies of this nucleus.

The neutron-rich $^{149}$Ce nucleus with N=91 neutrons lies in a transitional region between nearly spherical lighter cerium isotopes and deformed $^{150}$Ce with R=$E_{4+}/E_{2+} \approx 3.15$ ratio[1] close to the rotational model value of 3.33. In N=91 nuclei Nilsson levels related to $2f_{7/2}$, $1h_{9/2}$ and $1i_{13/2}$ orbitals are located close to each other and to the Fermi surface at quadrupole deformation parameter of about $\beta_2 \approx 0.2$. Consequently, the rotational bands based on the above orbitals are expected to occur at low excitation energies. Nilsson model calculations indicate, that the $2f_{7/2}$ and $1h_{9/2}$ orbitals are strongly mixed in this deformation region. As a result of the mixing, the calculated decoupling parameters for $1/2^-[541]$ state, $a_{541}$=0.65, and for

---

$1/2^-[530]$ state, $a_{530}=0.15$, are very different from values of about 4.6 and $-3.4$, respectively, expected for the pure states. With such low values of decoupling parameters the Coriolis mixing calculations performed with the use of the CORIOLIS code[2] show that all negative-parity rotational bands close to the Fermi level are normal strongly coupled bands with $\Delta I=1$ sequence and exhibit only small staggering of rotational levels. On the other side, for the positive-parity rotational bands close to the Fermi levels for N=91, the Coriolis mixing calculations predict existence of the decoupled rotational bands with $\Delta I=2$ sequence of spins.

To check these simple model predictions, an extensive total Routhian surface (TRS) calculations were performed for the $^{149}$Ce nucleus with the aim to compare theoretical predictions with the results of experimental studies of the high spin states in this nucleus[3, 4] and the recent studies of the $^{149}$La $\beta$-decay carried out at the R2-0 reactor in Studsvik using the OSIRIS on-line fission product mass separator[5]. In the latter study tentative assignment of spin and parity for several low-spin levels has been proposed and some ambiguities in the interpretation of the level scheme of $^{149}$Ce were resolved allowing for a comparison with theoretical predictions.

A recently developed pairing-deformation self-consistent total Routhian surface (SC_TRS) model[6, 7] has been used. Within this model, the total Routhian of the nucleus (energy in the rotating frame) is calculated as a sum of macroscopic part and a microscopic correction accounting for shell effects and pair correlation. The liquid drop model of Ref.[8] and the shell correction method of Strutinsky[9] were employed. The total Routhian can be written as

$$E^\omega(Z,N,\hat{\beta}) = E^{\omega=0}(Z,N,\hat{\beta}) + \left\{\hat{H}^\omega(Z,N,\hat{\beta}) - \hat{H}^{\omega=0}(Z,N,\hat{\beta})\right\} \qquad (1)$$

where $E^{\omega=0}(Z,N,\hat{\beta})$ represents the liquid drop energy, the single particle shell corrections, and pairing energy at frequency zero. The term in the brackets of Eq. (1) describes the change in energy due to rotation and $\hat{\beta}$ denotes deformation parameters $\beta_2$, $\beta_4$ and $\gamma$. The total Routhian is calculated on a grid in deformation space, including quadrupole deformation ($\beta_2$), triaxiality ($\gamma$) and hexadecapole ($\beta_4$) degrees of freedom, and then minimized with respect to the shape parameters to obtain equilibrium parameters. The deformed Woods-Saxon potential with an improved universal parameters set was used as a single-particle potential. In addition to the standard self-consistency requirement for the shape degrees of freedom, in this model the pairing correlations are also treated self-consistently. For more detailed description of the used TRS model see Ref.[10].

The calculated equilibrium deformation parameters of the positive-parity rotational band at $\hbar\omega \approx 0.075$ MeV lie around the values of $\beta_2 = 0.228$, $\gamma = -0.2°$ and $\beta_4 = 0.085$ while for the negative-parity rotational band these parameters are close to the values of $\beta_2 = 0.224$, $\gamma = 0.1°$ and $\beta_4 = 0.080$. Both total Routhian surfaces minima are well defined and show no indication of $\gamma$-softness. They remain pretty stable with increasing rotational frequency up to $\hbar\omega \approx 0.5$ MeV. The main components of the wave functions of the lowest positive and negative parity bands are, respectively, the $3/2^+[651]$ and $3/2^-[521]$ configurations, in agreement with the Coriolis mixing calculations[5]. Relatively large calculated values of the hexadecapole deformation parameter, especially for the positive-parity band, and, simultaneously, low values of the $\gamma$ non-axial deformation are worth to note.

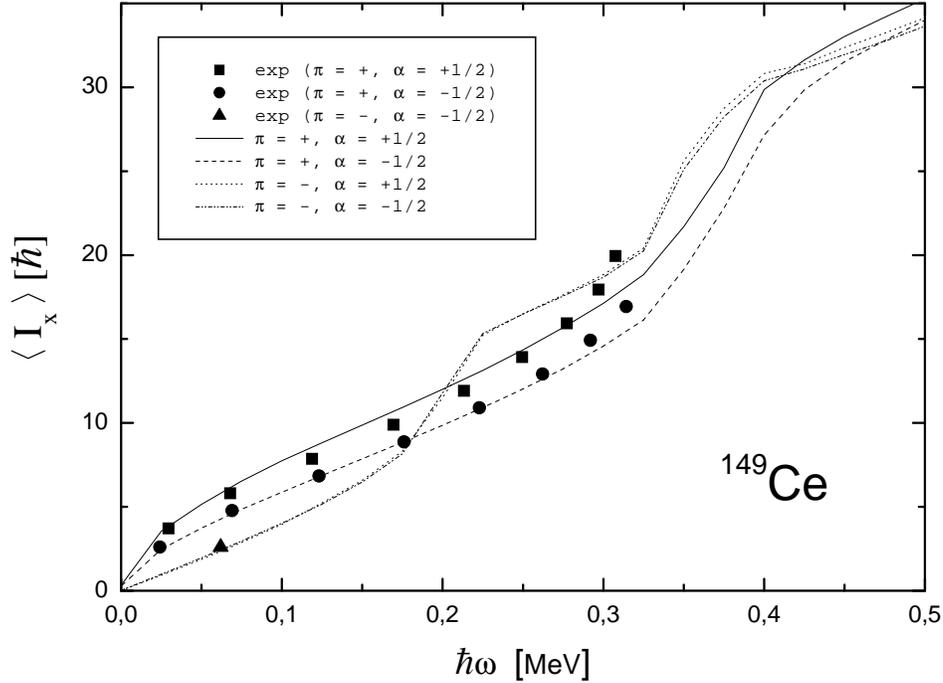

Fig. 1. The calculated and experimental values of the total angular momentum as a function of rotational frequency for both signatures of the lowest positive- and negative-parity rotational bands in $^{149}$Ce.

In order to check the possibility that the obtained large values of the hexadecapole deformation parameter are generated due to neglecting higher order multipole expansion terms, e.g. $\beta_6$, in our TRS model calculations, an additional calculations were performed minimizing potential energy of the lowest positive and negative states as a function of $\beta_2$, $\beta_3$, $\beta_4$, $\beta_5$ and $\beta_6$ deformation parameters at $\hbar\omega = 0$ MeV. The resulting $\beta_2$ and $\beta_4$ parameters did not change significantly in the presence of higher multipole terms. Other calculated deformation parameters were very close or equal to zero, showing in particular that at low values of rotational frequency the octupole degree of freedom plays negligible role for the lowest states in the $^{149}$Ce nucleus. Only the $5/2^+[642]$ configuration, which becomes dominant in the positive parity rotational band wave function at higher rotational frequencies ($\hbar\omega > 0.3$ MeV), shows significant octupole deformation of $\beta_3 \approx 0.10$ with $\beta_5 \approx 0.03$.

The calculated values of the angular momentum $<I_x>$ as a function of rotational frequency $\hbar\omega$ are shown in Fig.1 for both positive- and negative-parity rotational bands. Similarly to simple Coriolis mixing calculations a sizeable signature splitting is predicted only for the positive parity band. The backbending for the negative parity band is predicted at much lower frequency of about 0.20 MeV than for the positive parity band (about 0.33 MeV).

The experimental data (full circles and squares) for the rotational band built on the $3/2^+$, 133.5 keV level with a new spin and parity assignments proposed in Ref.[5] match very well theoretical predictions only if the high spin structure represents the positive parity band. Consequently, it excludes the possibility that the rotational band observed in experiment has a negative parity. Also the backbending frequency observed in the experiment (0.30 MeV), agrees only with prediction for the positive-parity band, although is about 0.03 MeV lower than the calculated value. Moreover, the only experimental point (full triangle) for the ground state rotational band also agrees very well with theoretical predictions for the negative parity band. The obtained good agreement supports the interpretation of the level scheme of $^{149}$Ce nucleus presented in Ref.[5].

**Acknowledgements**

The authors are indebted to Drs W. Satuła and R. Wyss for providing us with the latest version of the TRS code. We wish also to thank Dr W. Satuła for fruitful discussion. This work was supported in part by the KBN grant, and the Swedish Natural Science Research Council.

REFERENCES

[1] R.B. Firestone, Table of Isotopes, $8^{th}$ Edition, 1996
[2] R. Kaczarowski, Comput. Phys. Commun. **13** (1977) 63
[3] B.R.S. Babu, S.J. Zhu, A.V. Ramayya, J.H. Hamilton, L.K. Peker, M.G. Wang, T.N. Ginter, J. Kormicki, W.C. Ma, J.D. Cole, R.Aryaeinejad, K. Butler-Moore, Y.X. Dardenne, M.W. Drigert, G.M. Ter-Akopian, Yu.Ts.

Oganessian, J.O. Rasmussen, S. Asztalos, I.Y. Lee, A.O. Macchiavelli, S.Y. Chu, K.E. Gregorich, M.F. Mohar, S.Prussin, M.A. Stoyer, R.W. Lougheed, K.J. Moody, and J.F. Wild, Phys. Rev. C **54**, 568 (1996)

[4] F. Hoellinger, N. Schulz, J.L. Durell, I. Ahmad, M. Bentaleb, M.A. Jones, M. Leddy, E. Lubkiewicz, L.R. Morss, W.R. Phillips, A.G. Smith, W. Urban, and B. J. Varley, Phys. Rev. C **56**, 1296 (1997)

[5] A. Syntfeld, H. Mach, R. Kaczarowski, W. Płóciennik, W. Kurcewicz, B. Fogelberg and P. Hoff, *to be published*

[6] W. Satuła, R. Wyss, and P. Magierski, Nucl. Phys. **A578**, 45 (1994)

[7] W. Satuła and R. Wyss, Phys. Rev. C **50**, 2888 (1994)

[8] S. Cohen, F. Plasil, and W.J. Światecki, Ann. Phys. (N.Y.) **82**, 557 (1974)

[9] V.M. Strutinsky, Yad. Fiz. **3**, 614 (1966); Nucl. Phys. **A95**, 420 (1967)

[10] K. Starosta, Ch. Droste, T. Morek, J. Srebrny, D. B. Fossan, D. R. LaFosse, H. Schnare, I. Thorslund, P. Vaska, M. P. Waring, W. Satuła, S. G. Rohoziński, R. Wyss, I. M. Hibbert, R. Wadsworth, K. Hauschild, C. W. Beausang, S. A. Forbes, P. J. Nolan, and E. S. Paul, Phys. Rev. C **53**, 137 (1996)